\documentclass[11pt,epsf,letterpaper]{article}%
\usepackage{subfig}
\usepackage{color}
\usepackage{amsmath}
\usepackage{amsfonts}
\usepackage{verbatim}
\usepackage{amssymb}
\usepackage{graphicx}
\usepackage{epstopdf}
\usepackage{mathrsfs}
\usepackage{epsfig}
\usepackage{cite}%
\providecommand{\U}[1]{\protect\rule{.1in}{.1in}}
\providecommand{\U}[1]{\protect\rule{.1in}{.1in}}

\def\pa{\partial}

\begin{document}

\title{\center{{\huge {Scalar charges and the first law of black hole thermodynamics}}}}
\author{Dumitru Astefanesei$^{(1)}$, Romina Ballesteros$^{(1)}$, David Choque$^{(2,3)}$ and Ra\'ul Rojas$^{(1)}$\\
	\\\textit{$^{(1)}$Pontificia Universidad Cat\'olica de
		Valpara\'\i so, Instituto de F\'\i sica,} \\\textit{Av. Brasil 2950, Valpara\'{\i}so, Chile.} \\
	\\\textit{$^{(2)}$Universidad  Adolfo Ib\'{a}\~{n}ez, Dept. de Ciencias,
		Facultad de Artes Liberales,} \\\textit{Av. Padre
		Hurtado 750, Vi\~{n}a del Mar, Chile.}\\
	\\\textit{$^{(3)}$Universidad Nacional de San Ant\'onio Abad del Cusco,} \\\textit{Av. La Cultura 733, Cusco, Per\'u.}\\[1.2mm]
	%\texttt{{\small dumitru.astefanesei@pucv.cl,romina.ballesteros.n@gmail.com,}}\\
	%\texttt{{\small brst1010123@gmail.com,raulox.2012@gmail.com}}
	}
\maketitle

\begin{abstract}
	We present a variational formulation of Einstein-Maxwell-dilaton theory in flat spacetime, when the asymptotic value of the scalar field is not fixed. We obtain the boundary terms that make the variational principle well posed and then compute the finite gravitational action and corresponding Brown-York stress tensor. We show that the total energy has a new contribution that depends of the asymptotic value of the scalar field and discuss the role of scalar charges for the first law of thermodynamics. We  also  extend  our analysis to hairy black holes in Anti-de Sitter spacetime and investigate the thermodynamics of an exact solution that breaks the conformal symmetry of the boundary. 
	\newline
	
\end{abstract}

\newpage

%%%%%%%%%%%%%%%%%%%%%%%%%%%%%%%%%%%%%%%%%%%%%%%%%%%%%%%%%%%%%%%%%%%%%%%%%%%%%%%%%%%%%%%%%%%%%%%%

\section{Introduction}

%%%%%%%%%%%%%%%%%%%%%%%%%%%%%%%%%%%%%%%%%%%%%%%%%%%%%%%%%%%%%%%%%%%%%%%%%%%%%%%%%%%%%%%%%%%%%%%%
Scalar fields play a central role in particle physics and cosmology and arise naturally
in the high energy physics unification theories. It is then important to understand  generic properties of gravity theories coupled to scalars (and other matter fields),  particularly the role of scalars for black hole physics.\footnote{Some recent interesting applications can be found in \cite{Hirschmann:2017psw,Jai-akson:2017ldo,Cardenas:2017chu,Julie:2017rpw,McCarthy:2018zze,Brihaye:2018woc}.} 

The low energy effective actions of string theory correspond to (consistent 
truncations of) gauged supergravities (see, e.g., \cite{Trigiante:2016mnt} and references therein) and some of the accepted wisdoms of general relativity 
may be reconsidered in this context. One of the important differences is that, contrary to fixing the boundary conditions as in general relativity, the boundary conditions in string theory are determined by dynamical vacuum expectation values (VEVs) of the moduli. An important and unusual consequence is that, for the non-extremal  black 
holes in string theory, both the mass  and area of the horizon depend in a 
non-trivial way on the asymptotic values of the moduli $\phi^a_{\infty}$ (here, $a$ labels different scalars), which leads to a drastic modification of the first law of (static) black hole thermodynamics \cite{Gibbons:1996af}:
\begin{equation}
\label{1stlaw}
dM = TdS + \Psi dQ + \Upsilon dP + \left(\frac{\partial M}{\partial \phi_{\infty}^a} \right) d\phi_{\infty}^a
\end{equation}
where $\Psi$ and $\Upsilon$ are the electric and magnetic potentials, and the coefficients of $\phi_{\infty}^a$ are computed at fixed charges and entropy,
\begin{equation}
\left(\frac{\partial M}{\partial \phi_{\infty}^a} \right)_{S,Q,P} = -G_{ab}(\phi_\infty)\Sigma^b
\end{equation}
With the notations of \cite{Gibbons:1996af}, $G_{ab}(\phi_\infty)$ is the metric of the moduli space and $\Sigma^a$ are the scalar charges that can be read of from the asymptotic expansion of the scalar field at the spatial infinity:
\begin{equation}
\phi^a = \phi_{\infty}^a + \frac{\Sigma^a}{r}+O(\frac{1}{r^2})
\end{equation}  
A similar proposal appears in the context of AdS/CFT duality where, for an exact hairy black hole solution that is asymptotically AdS, it was found that the first law should be modified by an additional conjugate pair $(X,Y)$ of thermodynamic variables \cite{Lu:2013ura}:
\begin{equation}
dM = TdS + \Psi dQ + \Upsilon dP + XdY
\end{equation}
These quantities, $(X,Y)$, are expressible as functions of the conserved charges $(M,P,Q)$ and were interpreted in their own right as a scalar charge and its conjugate potential \cite{Lu:2013ura}.

One problem with the first law of thermodynamics (\ref{1stlaw}) for stringy black holes is that the scalar charges are not conserved charges. They correspond to degrees of freedom living outside the horizon (the `hair') and are not associated to a new independent integration constant, that is why it is called `secondary hair'. In string theory, the scalar fields (moduli) are interpreted as local coupling constants and so a variation of their boundary values is equivalent to changing the couplings  of  the  theory. A resolution was proposed in \cite{Astefanesei:2006sy} (see, also, \cite{Astefanesei:2009wi}):  one can in principle redefine the charges so that the mass and scalar charges do not depend of $\phi_{\infty}$, but the price to pay is that the new `dressed electric and magnetic charges' are not the physical ones. If the asymptotic value of the scalar field is non-zero, but fixed  directly in the action, $\phi_\infty= const.$, that corresponds to a different theory with a different coupling (the factor $e^{\phi_\infty}$ is absored in the coupling constant not in the values of the charges) for the gauge field and, within that theory, the term $\Sigma d\phi_{\infty}$ vanishes. This proposal was made concrete in \cite{Hajian:2016iyp} where, by using the `phase space method', it was shown that this is a valid integrability condition and there is no need for an extra contribution of the scalar field in the first law. 

However, we would like to emphasize that the proposal of  Gibbons, Kallosh, and Kol \cite{Gibbons:1996af} is about a variation of the boundary conditions for the scalar fields and so, despite the arguments in \cite{Astefanesei:2006sy, Hajian:2016iyp}, it stays robust and intriguing. The main question, which still remains, is then why the scalar charges that act as sources for the moduli, but are not conserved, appear in the first law of black hole thermodynamics when considering variations of $\phi_\infty$?

In this work, we investigate the role of non-trivial boundary conditions of the scalar field in Einstein-Maxwell-dilaton theory. We are interested in asymptotically flat hairy black hole  solutions for which the asymptotic value of the scalar can vary and asymptotically AdS dyonic hairy black hole solutions for which the scalar breaks the conformal symmetry of the boundary. In flat spacetime, we obtain a well-posed variational principle by adding a new boundary term to the action, which permits us to compute the correct total energy, and clarify the role of the (non-conserved) scalar charges to the first law of thermodynamics \cite{Gibbons:1996af}. Armed with the intuition from flat spacetime, we show that once the energy is also correctly obtained in AdS spacetime \cite{Anabalon:2014fla}, when the boundary conditions of the scalar field do not preserve the isometries of AdS at the boundary \cite{Henneaux:2006hk}, the first law is satisfied and there is no need to consider an extra contribution from the scalar field. These considerations are of special interest when considering the embedding in string theory and the scalar field (dilaton) becomes dynamical and for AdS holographic applications, e.g. the hairy black holes can be used to describe symmetry breaking or phase transitions in the dual quantum field theory.

\section{Asymptotically flat hairy black holes}
In this section, we propose a variational principle for asymptotically flat hairy black holes when the boundary values of the scalar fields can vary and show that the total energy has a new contribution that is relevant for thermodynamics. The goal of this section is to discuss this issue concretely in the simplest possible non-trivial setting, namely we are going to use the quasilocal formalism of Brown and York \cite{Brown:1992br} for a theory with only one scalar field that is coupled to the gauge field.

\subsection{The first law of hairy black hole thermodynamics}
We start with a  brief review of \cite{Gibbons:1996af} and, for clarity, we explicitly obtain the scalar charge term in the first law for an exact hairy black hole solution.
Besides the graviton, every string theory contains another universal state, a massless scalar called the dilaton $\phi$. We consider the Einstein-Maxwell-dilaton action 
\begin{equation}
I\left[g_{\mu\nu},A_\mu,\phi\right]= I_{bulk} + I_{GH} =
\frac{1}{2\kappa}\int_{\mathcal{M}}
{d^4x\sqrt{-g}
	\left(R-
	e^{\alpha\phi}F_{\mu\nu}F^{\mu\nu}
	-2\pa_\mu\phi\pa^\mu\phi\right)}
+\frac{1}{\kappa}\int_{\pa\mathcal{M}}
{d^3x\sqrt{-h}K}
\label{action0}
\end{equation}
where $\kappa=8\pi G_N$ and, with our conventions, $G_N=1$.  The second term is the Gibbons-Hawking boundary term and $K$ is the trace of the extrinsec curvature $K_{ab}$ defined on the boundary $\pa\mathcal{M}$ with the induced metric $h_{ab}$.

The coupling between the scalar field and gauge field in the action (\ref{action0}) appears in the low energy actions of string theory for particular values of $\alpha$, though in our analysis we are going to keep $\alpha$ arbitrary. The equations of motion for the metric, scalar, and gauge field are
\begin{align}
E_{\mu\nu}:=R_{\mu\nu}
-2\pa_\mu\phi\pa_\nu\phi
-2\text{e}^{\alpha\phi}
\left(F_{\mu\alpha}F_{\nu}{}^{\alpha}
-\frac{1}{4}g_{\mu\nu}
F_{\alpha\beta}F^{\alpha\beta}\right)&=0 \\
\frac{1}{\sqrt{-g}}\pa_\mu\left(\sqrt{-g}
g^{\mu\nu}\pa_\nu\phi\right)
-\frac{1}{4}\alpha \text{e}^{\alpha\phi}F_{\mu\nu}F^{\mu\nu}&=0\\
\pa_\mu
\left(\sqrt{-g}\text{e}^{\alpha\phi}
F^{\mu\nu}\right)&=0 \label{maxw}
\end{align}
The general metric ansatz for a static dyonic hairy black hole solution  is 
\begin{equation}
ds^{2}=-a^2dt^2+a^{-2}dr^2+b^2(d\theta^{2}+\sin^{2}{\theta}\, d\varphi^{2})
\end{equation}
where the metric functions have only radial dependence, $a=a(r)$ and $b=b(r)$. The gauge field compatible with this ansatz and the equations of motion is
\begin{equation}
\label{gauge}
F = -\frac{qe^{-\alpha\phi}}{b^2} dt\wedge dr
-p\sin\theta\, d\theta\wedge d\varphi 
\end{equation}
The combination $E_{t}^{t}+E_{\theta}^{\theta}$ leads to an integrable equation $(a^2b^2)''=2$ (where the prime symbol means derivative with respect to $r$) with the general solution
\begin{align}
a^{2}=\frac{(r-r_{+})(r-r_{-})}{b^2}
\end{align}
Since we are interested in asymptotically flat solutions, we consider the boundary expansion $a^2=1+O(r^{-1})$ that fixes the fall-off of the other metric function as 
\begin{equation}
b^2=r^2+\beta r+\gamma+O(r^{-1})
\end{equation}
By using this expressions into the combination $E_{t}^{t}-E_{r}^{r}=0$, which leads to the equation $b''+b\phi'^2=0$, we obtain the following asymptotic form of the scalar field
\begin{equation}
\phi=\phi_{\infty}+\frac{\Sigma}{r}+O(r^{-2})
\label{scalar}
\end{equation}
where $\phi_{\infty}$ is the boundary condition for the theory we have considered (for the scalar field) and $\Sigma$ is the scalar charge. One can easily obtain that $4\Sigma^2=\beta^2-4\gamma$ and for  $\beta=0$ we have $b^2=r^2-\Sigma^2$, which is indeed the case for the exact solution in the theory with $\alpha=-2$ that will be presented below.

To concretely check the steps for obtaining (\ref{1stlaw}), we are going to use an exact four dimensional hairy black hole solution \cite{Gibbons:1987ps}, when the magnetic charge vanishes  and the scalar is coupled with the exponential parameter $\alpha=-2$. The equations of motion can be analitically solved for any value of $\alpha$ (see, \cite{Garfinkle:1990qj}) and the exact black hole solution is
\begin{equation}
a^2=\frac{(r-r_+)(r-r_-)}{r^2-\Sigma^2} \,,\qquad
b^2=r^2-\Sigma^2 
\,,\qquad
\phi=\phi_\infty
+\frac{1}{2}\ln\left(\frac{r+\Sigma}{r-\Sigma}\right)
\end{equation}
where 
\begin{equation}
r_-=-\Sigma,\qquad r_+=\Sigma
-\frac{(qe^{\phi_{\infty}})^2}{\Sigma}
\end{equation}

The ADM mass \cite{Arnowitt:1959ah,Arnowitt:1960es} is obtained by expanding the $g_{tt}$ component of the metric,
\begin{equation}
-g_{tt}=a^2
=\frac{\Sigma r+(qe^{\phi_\infty})^2-\Sigma^2}
{\Sigma(r-\Sigma)}
=1+\frac{\left(qe^{\phi_\infty}\right)^2}{\Sigma{r}}
+O\left(r^{-2}\right) 
\end{equation}
which leads to the following expression:
\begin{equation}
\label{mass}
M=-\frac{\left(qe^{\phi_\infty}\right)^2}{2\Sigma}
\end{equation}
with the scalar charge negative, $\Sigma < 0$. The physical (conserved) electric charge $Q$ is computed, as usual, by integrating the equation of motion for the electric field and, with our conventions, the result is
\begin{equation}
Q = \frac{2}{\kappa}\oint{e^{-2\phi}\star{F}}=\frac{2}{\kappa}\oint{e^{-2\phi}
	\left(\frac{1}{4}\sqrt{-g}\epsilon_{\alpha\beta\mu\nu}
	F^{\alpha\beta}dx^\mu\wedge dx^\nu\right)} =q
\end{equation}

At this point, it is important to observe that, when evaluating on an exact solution, $\Sigma$ is not an independent integration constant and the solution is regular\footnote{When both (electric and magnetic) charges are non-zero, there exist two horizons. However, in this special case with only one non-zero electric field, there exists only one horizon $r=r_{+}$ because $r=r_{-}$ corresponds to a real singularity. The regularity condition $r_{+} >r_{-}$ is, from a physical point of view, equivalent with the fact that there is a maximum charge that can be carried by the black hole.} only if $2M^2-Q^2e^{2\phi_\infty}> 0$. With the mass (\ref{mass}), one can explicitly check the first law (\ref{1stlaw}), which contains the extra term proposed in \cite{Gibbons:1996af}, and  Smarr formula 
\begin{equation}
M=2TS+Q\Psi
\label{smarr}
\end{equation}
that, despite the extra term in the first law, does not contain the $\Sigma\phi_\infty$ contribution.  

\subsection{Total energy and Brown-York formalism}
There are by now examples \cite{Astefanesei:2009wi, Astefanesei:2005ad, Mann:2005yr, Mann:2006bd, Astefanesei:2006zd, Astefanesei:2009mc, Astefanesei:2010bm, Herdeiro:2010aq, Compere:2011db, Compere:2011ve} of the first law of thermodynamics and conserved charges obtained by using the quasilocal formalism of Brown and York \cite{Brown:1992br} supplemented with counterterms for asymptotically flat spacetimes \cite{Lau:1999dp,Mann:1999pc,Kraus:1999di,Mann:2005yr}. For Einstein-Maxwell theory in four dimensions, the action (\ref{action0}) should be supplemented with a counterterm that cancel the IR divergencies,
\begin{equation}
I=I_{bulk}+I_{GH}+I_{ct} \,\,,\qquad I_{ct}=-\frac{1}{\kappa}\int_{\pa\mathcal{M}}
{d^3x\sqrt{-h}\sqrt{2\mathcal{R}^{(3)}}}
\end{equation}
where $\mathcal{R}^{(3)}$ is the Ricci scalar of the $3$-dimensional boundary metric $h_{ab}$. In the presence of a magnetic field, one should also carefully consider the variational principle for the gauge field \cite{Hawking:1995ap}. 

For simplicity, we start with the boundary condition when $\Sigma$ is constant in the fall-off (\ref{scalar}). However,  at the end of this section we present the boundary term for a general boundary condition of the form $\Sigma(\phi_\infty)$ that leads again to the same physical interpretation. For Einstein-Maxwell-dilaton theory, due to the variation of the moduli, to obtain a well defined variational principle when the scalar charge $\Sigma$ is kept fixed, we have to add a new boundary term:
\begin{equation}
\label{scalarct}
I_\phi=-\frac{2}{\kappa}\int_{\pa\mathcal{M}}
{d^3x\sqrt{-h}\left[
	\frac{\phi_\infty}{\Sigma}(\phi-\phi_\infty)^2
	\right]}
\end{equation}

To get the free energy $F$ of a hairy black hole, we shall compute the on-shell action on the Euclidean section:
\begin{equation}
I^E =\beta F =\beta\left(M-TS-Q\Psi-P\Upsilon+\Sigma\phi_\infty\right)
\end{equation}
where the periodicity of the Euclidean time is related to the black hole temperature by 
$\beta = 1/T$. 

We observe that there is an extra term $\Sigma\phi_\infty$ that is, in fact, coming from the scalar field counterterm $I_{\phi}$, though, as we already saw, a similar term does not appear in the Smarr formula (\ref{smarr}). This is an important hint that a computation of the total energy could be different from the $ADM$ mass when considering the boundary term $I_{\phi}$. With all the terms required for a correct variational principle, we obtain the regularized quasilocal stress tensor of \cite{Astefanesei:2005ad}, but now supplemented with the contribution from the scalar field
\begin{equation}
\tau_{ab}=\frac{1}{\kappa}\left[
K_{ab}-h_{ab}K-\Phi(\mathcal{R}^{(3)}_{ab}-h_{ab}\mathcal{R}^{(3)})-h_{ab}\Box\Phi+\Phi_{;ab}\right]
+\frac{2h_{ab}}{\kappa}\left[\frac{\phi_{\infty}}{\Sigma}(\phi-\phi_{\infty})^2\right]
\end{equation}
where
\begin{equation}
\Phi=\sqrt\frac{2}{\mathcal{R}^{(3)}}
\end{equation}
If we consider a collection of observers on the boundary of a static black hole spacetime, since $\xi^\mu=\delta^\mu_t$ is a Killing vector, they are going to measure the same (total) energy that is the conserved charge associated with this specific Killing vector defined as \cite{Brown:1992br}:
\begin{equation}
\label{BYcharge}
E=\oint_{\Xi}{d^2\Xi\sqrt{\sigma}n^a\xi^b\tau_{ab}}
\end{equation}
Here, $\Xi$ is a two dimensional closed surface with the unit normal $n^a$ and the induced metric
\begin{equation}
\sigma_{ij}dx^idx^j=b^2(d\theta^2+\sin^2\theta d\varphi^2)
\end{equation}
Since we are interested in static configurations, there are no gravitational waves and so there is no need to consider the null infinity in our analysis. Evaluating this conserved quantity, (\ref{BYcharge}), at the spatial infinity, we obtain the following expression for the total energy:
\begin{equation}
E_{\text{total}}=M+\phi_{\infty}\Sigma
\end{equation}
This result was obtained for a fixed scalar charge, $\Sigma=constant$, and this leads to the following first law of thermodynamics
\begin{equation}
dE_{\text{total}} = T dS + \Psi dQ + \Upsilon dP
\end{equation}
with the $\Sigma d\phi_\infty$ reabsorbed in the total energy of the spacetime, which is different from the ADM mass.

We should now consider a more general boundary condition of the form 
\begin{equation}
\Sigma\equiv \frac{dW(\phi_\infty)}{d\phi_\infty}
\end{equation}
which is very similar with the one proposed in \cite{Hertog:2004ns} for the hairy black holes in AdS. When evaluating on an exact solution, this is an integrability condition for the Hamiltonian mass, which is equivalent to the one for AdS (see the eq. $(17)$ of \cite{Anabalon:2014fla} and the comments there) . The general boundary term is now
\begin{equation}
\tilde I_{ct}^\phi=-\frac{2}{\kappa}\int_{\pa\mathcal{M}}
{d^3x\sqrt{-h}\left[\frac{(\phi-\phi_\infty)^2}
{\Sigma^2}W(\phi_\infty)\right]}
\end{equation}
and matches (\ref{scalarct}) when $\Sigma$ is constant. As expected, a similar computation of the total energy yields the following result:
\begin{equation}
E_{total} = M + W(\phi_\infty)
\end{equation}
where $M$ is the ADM mass obtained from the expansion of $g_{tt}$ at spatial infinity.

\section{Asymptotically AdS hairy black holes}
In this section, we compute the energy of an exact dyonic hairy black hole proposed in \cite{Lu:2013ura} and check the first law of thermodynamics.\footnote{In the past years, there were constructed exact regular hairy black hole solutions \cite{Acena:2013jya,Anabalon:2013qua,Anabalon:2013sra,Acena:2012mr,Anabalon:2013eaa, Anabalon:2016izw} for a specific moduli potential, which finally was shown to correspond to extended supergravity models \cite{Anabalon:2017yhv,DallAgata:2012mfj} and our analysis can be also applied to this case.}  We follow closely \cite{Anabalon:2014fla} because it is technically easier from a practical point of view\footnote{The counterterm method in AdS was developed in \cite{Balasubramanian:1999re,Skenderis:2000in,Henningson:1998gx} and in the presence of scalar fields with mixed boundary conditions in \cite{Papadimitriou:2007sj, Anabalon:2015xvl,Marolf:2006nd}.} and the energy can be read of directly from the metric even if the trace anomaly does not vanish. We are going to show that, even when the boundary conditions for the scalar field breaks the conformal invariance, the first law is again satisfied without introducing the extra terms depending of the scalar charges. 

\subsection{Dyonic hairy black hole}
We consider the theory described by the action \cite{Lu:2013ura}
\begin{equation}
I=\frac{1}{2\kappa}\int_{\mathcal{M}}
{d^4x\sqrt{-g}
	\left[
	R-\frac{1}{2}(\pa\phi)^2
	-\frac{1}{4}e^{-\sqrt3\phi}F^2
	+\frac{6}{l^2}
	\cosh\left(\frac{1}{\sqrt3}\phi\right)
	\right]}
\end{equation}
and the following regular static dyonic hairy black hole,
\begin{align}
ds^2 &= 
-(H_1 H_2)^{-\frac{1}{2}}f dt^2
+(H_1 H_2)^{\frac{1}{2}}
\left[\frac{dr^2}{f}
+r^2\left(d\theta^2+\sin^2\theta\, d\varphi^2\right)\right] 
\label{lupopemetric}\\
\phi &=\frac{\sqrt{3}}{2} \ln\left(\frac{H_2}{H_1}\right) \\
A_{\mu} dx^\mu &=\sqrt2 \left(\frac{1-\beta_1 f_0}
{\sqrt{\beta_1\gamma_2}\, H_1}\, dt + 2\mu\,\gamma_2^{-1}\sqrt{\beta_2\gamma_1}\, \cos\theta\, d\varphi\right)
\label{lupopegauge}
\end{align}
where the relevant functions were obtained in \cite{Lu:2013ura}:
\begin{align}
f&=f_0 + \frac{r^2}{l^2} H_1 H_2\,,\qquad f_0=1 - \frac{2\mu}{r} \\
H_1&=\gamma_1^{-1} (1-2\beta_1 f_0 + \beta_1\beta_2 f_0^2)\,,\qquad
H_2=\gamma_2^{-1}(1 - 2\beta_2 f_0 + \beta_1\beta_2 f_0^2) \\
\gamma_1&= 1- 2\beta_1 + \beta_1\beta_2\,,\qquad \gamma_2 = 1-2\beta_2 + \beta_1\beta_2\,.
\end{align}

We would like to verify the first law for this specific solution and explicitly check if the scalar charge term is relevant for the termodynamics in this case \cite{Lu:2013ura}. The electric and magnetic conserved charges can be easily obtained
\begin{equation}
Q=\frac{\mu\sqrt{\beta_{1}\gamma_{2}}}{\gamma_{1}\sqrt{2}} \,, \qquad P=\frac{\mu\sqrt{\beta_{2}\gamma_{1}}}{\gamma_{2}\sqrt{2}}
\end{equation}
and their conjugate potentials are 
\begin{equation}
\Psi=\sqrt{\frac{2}{\beta_1\gamma_2}}
\left[
1-\beta_1-\frac{1-\beta_1f_0(r_+)}{H_1(r_+)}
\right] \,, \qquad \Upsilon=\sqrt{\frac{2}{\beta_2\gamma_1}}
\left[
1-\beta_2-\frac{1-\beta_2f_0(r_+)}{H_2(r_+)}
\right]
\end{equation} 
To compute the mass as in \cite{Lu:2013ura}, we should use canonical coordinates for which the factor in front of the angular part of the metric becomes $b^2 = \rho^2 + O(\rho^{-1})$ in the asymptotic limit:
\begin{equation}
r=\rho+c_1+\frac{c_2}{\rho}+O(\rho^{-2})
\end{equation}
where 
\begin{equation}
c_1=\frac{\mu(2\beta_1^2\beta_2^2-3\beta_1^2\beta_2-3\beta_1\beta_2^2+6\beta_1\beta_2-\beta_1-\beta_2)}
{\gamma_1\gamma_2} \,, \qquad c_2=\frac
{3\mu^2(1-\beta_1\beta_2)^2(\beta_1-\beta_2)^2}
{2\gamma_1^2\gamma_2^2}
\end{equation}
With this change of coordinates, we obtain the following fall-off for the metric function:
\begin{equation}
-g_{tt}= 1 -\frac{2M}{\rho}+ \frac{\rho^2}{l^2} = 1+ \frac{\rho^2}{l^2}-\frac{2(1-\beta_1)(1-\beta_2)(1-\beta_1\beta_2)\mu}{\gamma_1\gamma_2\,\rho}+O(\rho^{-2})
\end{equation}
The first law is satisfied only if an extra term, $XdY$, is added \cite{Lu:2013ura}
\begin{equation}
dM=TdS+\Psi dQ+\Upsilon dP+XdY
\label{first1}
\end{equation}
where
\begin{equation}
X=\frac{4\mu^3(\beta_1-\beta_2)
	\sqrt{\beta_1\beta_2^3}}
{l^2(1-\beta_1\beta_2)\gamma_2^2},
\qquad
Y=\frac{\sqrt{\beta_1}\gamma_2}
{\sqrt{\beta_2}\gamma_1}
\end{equation}
%

%%%%%%%%%%%%%%%
\subsection{Boundary conditions and trace anomaly}
As in the asymptotically flat case, we would like to understand if the term $XdY$ can be reabsorbed in a correct definition of the total energy.\footnote{It is interesting to observe that by keeping $X$ constant for this particular solution, one obtains an integrability condition similar with $\Sigma$ constant for flat space.} We follow closely \cite{Anabalon:2014fla, Anabalon:2015xvl} and first check that, indeed, the boundary conditions for the scalar field break the conformal invariance \cite{Henneaux:2006hk}. The fall-off of the scalar field is
\begin{equation}
\phi(\rho)=\frac{A}{\rho}+\frac{B}{\rho^2}
+O(\rho^{-3})
\end{equation}
where 
\begin{align}
A(\mu,\beta_1,\beta_2)&=\frac
{2\sqrt{3}\mu(\beta_2-\beta_1)(1-\beta_1\beta_2)}
{\gamma_1\gamma_2} \\
B(\mu,\beta_1,\beta_2)
&=\frac{2\sqrt{3}\mu^2(\beta_1-\beta_2)
	(\beta_1^3\beta_2^2+\beta_1^2\beta_2^3
	-8\beta_1^2\beta_2^2+6\beta_1^2\beta_2
	+6\beta_1\beta_2^2-8\beta_1\beta_2+\beta_1+\beta_2)}
{\gamma_1^2\gamma_2^2}
\end{align}
When $\beta_1=0$ or $\beta_2=0$, the coefficients $A$ and $B$ satisfy the conformal invariance condition $B\sim A^2$ and the solutions can be interpreted as triple-trace deformations in the dual field theory. However, in general, when both electric and magnetic charges are non-zero, the conformal symmetry is broken. Therefore, one also expects that this non-trivial behaviour of the scalar field with a general relation between the coefficients $A$ and $B$ affects directly the thermodynamics. Concretely, when (asymptotically) the isometries of AdS are preserved, the term $XdY$ vanishes and so, as expected, the extra contribution of the scalar field to the first law should be related to the trace anomaly. The Hamiltonian mass can be read of directly from the $g_{\rho\rho}$ (and not from $g_{tt}$) and it has a new contribution due to the back-reaction of the scalar field in the boundary (see the eqs. (12) and (18) of \cite{Anabalon:2014fla}):
\begin{equation}
g_{\rho\rho}=\frac{l^2}{\rho^2}
+\frac{Cl^4}{\rho^4}
+\frac{Dl^5}{\rho^5}
+O(\rho^{-6})
\end{equation}
where the coefficients $C=C(\mu,\beta_1,\beta_2)$ and $D=D(\mu,\beta_1,\beta_2)$ are given by:
\begin{align}
C&=-1-\frac
{3\mu^2(\beta_2-\beta_1)^2
	(1-\beta_1\beta_2)^2}
{l^2\gamma_1^2\gamma_2^2} \label{metric-a} \\
D&=\frac
{2\mu(1-\beta_1)(1-\beta_2)(1-\beta_1\beta_2)}
{l\gamma_1\gamma_2} \notag \\
&+\frac{8\mu^3(\beta_2-\beta_1)^2
	(1-\beta_1\beta_2)
	(\beta_1^3\beta_2^2+\beta_1^2\beta_2^3
	-8\beta_1^2\beta_2^2+6\beta_1^2\beta_2
	+6\beta_1\beta_2^2-8\beta_1\beta_2+\beta_1+\beta_2)}
{l^3\gamma_1^3\gamma_2^3}
\label{metric-b}
\end{align}
As a double check, it is important to emphasize that the relations   
\begin{align}
C&=-1-\frac{A^2}{4l^2} \\
D&=\frac{2M}{l}-\frac{2AB}{3l^3}
\end{align}
are satisfied and so, indeed, we can use the general result for the mass presented in \cite{Anabalon:2014fla, Anabalon:2015xvl}:
\begin{equation}
\label{EtAdS}
E_{\text{total}} = \frac{(1-\beta_1)(1-\beta_2)(1-\beta_1\beta_2)\mu}
{\gamma_1\gamma_2} + \frac{1}{4l^2}
\left(W-\frac{1}{3}AB\right) = M+\frac{1}{4l^2}
\left(W-\frac{1}{3}AB\right)
\end{equation}
where $B= \frac{dW}{dA}$. Notice that the conventions for the action in \cite{Anabalon:2015xvl} are slightly different from ours and, to match the results, one should rescale the scalar field accordingly. 

It is a straightforward computation to show that the term $XdY$ matches the variation of the extra term in (\ref{EtAdS}). Once again, by using the correct total energy, the first law of thermodynamics can be rewritten as
\begin{equation}
dM=TdS+\Psi dQ+\Upsilon dP+XdY \quad
\Leftrightarrow \quad
dE_{\text{total}}=TdS+\Psi dQ+\Upsilon dP
\end{equation}
and there is no need to add an extra contribution depending of the scalar charges.
\section{Conclusions}
In this paper, using ideas arising from the quasilocal formulation of energy and counterterm method, 
we have revised the first law of hairy black hole thermodynamics and shown that the (non-conserved) scalar charges can not appear as independent terms, neither in flat space nor in AdS.  The work of \cite{Anabalon:2017eri,Anabalon:2016yfg} in which it was proven that there is no independent integration constant associated with the scalar field for hairy black holes in AdS supports our conclusion for the AdS dyonic hairy black hole.\footnote{For other attempts to explain the extra term $XdY$ in the first law of thermodynamics of the AdS dyonic hairy black hole solution presented in \cite{Lu:2013ura}, see \cite{Anabalon:2016ece,Cardenas:2016uzx}. }

In string theory, there exists a dimensionless parameter $g_s$ (the string 
coupling) that is controlled by the VEV of the dilaton, $g_s= e^{<\phi>}$, which is not fixed by the equations of motion. In fact, string theory has no free parameters because all the coupling constants are fixed by moduli VEVs. Therefore, the values of moduli at infinity, $\phi_{\infty}$, can be interpreted as labelling a continous familiy of vacua of the theory. Changing the asymptotic values of the moduli is similarly with changing the coupling constants of the theory and so the same hairy black hole configuration can be interpreted in different theories. This is unusual in general relativity because the boundary conditions are fixed, but quite common in string theory. For example, to compute the black hole entropy of a supersymmetric extremal black holes in string theory, one is doing a D-brane computation in the weak coupling regime and, since this result is protected by supersymmetry, it remains the same in the strong coupling regime in which the black hole exists. From the point of view of general relativity, we have shown that a variation of $\phi_{\infty}$, when keeping the scalar charge fixed, is producing a new contribution to the total energy of the system and the first law of thermodynamics is satisfied without the need of including the scalar charge contribution. 

In AdS spacetime, a similar analysis can be done, though the interpretations are very different. Our main observation is that the boundary conditions for the dyonic hairy black hole solution in \cite{Lu:2013ura} break the conformal invariance and the correct energy should contain an extra contribution (due to the trace anomaly), which was calculated in \cite{Anabalon:2014fla,Anabalon:2015xvl} (see, also, \cite{Hertog:2004ns}). With this new expression for the energy, the scalar charges do not contribute and the usual first law of black hole thermodynamics that contains only the conserved charges is again satisfied for hairy solutions.    

We would also like to comment on the attractor mechanism \cite{Ferrara:1995ih} in this context: for the hairy extremal black holes, the horizon values of the moduli are fixed by the electric and magnetic charges, so the entropy. First, it is important to emphasize that the interpretations of the attractor mechanism in flat space and AdS are different. In flat space, even for extremal non-supersymmetric black holes \cite{Goldstein:2005hq, Sen:2005wa,  Astefanesei:2006dd}, the entropy does not depend of the asymptotic value of the moduli and so it is invariant with respect to the variation of the couplings. This simple observation is at the basis of the exact computation of the extremal black hole microscopic entropy in string theory without supersymmetry \cite{Dabholkar:2006tb, Astefanesei:2006sy}. However, in AdS spacetime, the moduli flow corresponds to an RG flow and the attractor mechanism in the dual field theory can be holographically interpreted as the fact that the IR physics does not depend about the UV details \cite{Astefanesei:2007vh, Astefanesei:2008wz, Astefanesei:2010dk}. Therefore, the attractor mechanism is about the variation of $\phi_{\infty}$, which is equivalent with a variation of coupling constants in flat space or an holographic RG flow in AdS and can be interpreted as a no-hair theorem for the extremal hairy black holes. Contrary to the claims of \cite{Hajian:2016iyp}, the attractor mechanism stays robust (just in the extremal case the entropy does not depend of the couplings) and there is no need to be revised.

\section*{Acknowledgments}
DA would like to thank Oscar Fuentealba for interesting conversations and Debora Rojas for her continuous support. We would also like to thank Marcela Cardenas, Felix-Louis Julie, and Nathalie Deruelle for useful correspondence. DA would like to thank CECs, Valdivia for the hospitality during the last stages of this research. This work has been funded by the Fondecyt Regular Grant 1161418 and Newton-Picarte Grant DPI20140115. The work of DC is supported by Fondecyt  Postdoc Grant 3180185. RR was also supported by the national Ph.D. scholarship Conicyt 21140024.

\end{document}